\documentclass[aps,pra,twocolumn,showpacs,groupedaddress]{revtex4-2}
\usepackage{graphicx}

\usepackage{color}
\usepackage{latexsym}

\usepackage{xcolor}
\usepackage{hyperref}
\usepackage{amsmath,amssymb}
\usepackage{float}
\usepackage{bm}

\begin{document}

\title{Spectral Singularities with Directional Sensitivity}
\author{Hamidreza Ramezani$^1$} 
\email {hamidreza.ramezani@utrgv.edu}

\affiliation{$^1$ Department of Physics and Astronomy, University of Texas Rio Grande Valley, Edinburg, TX 78539, USA }

\begin{abstract}
    We propose a class of spectral singularities that are sensitive to the direction of excitation and are arising in nonlinear systems with broken parity symmetry. These spectral singularities are sensitive to the direction of the incident beam and result in diverging transmission and reflection for the left (right) incident while the transmission and reflection of the right (left) side of the system remain finite. For pedagogical reason first, we review the scattering formalism in nonlinear systems using an abstract $\delta$-function model. Then using a parity symmetry broken nonlinear system consists of two delta functions, one linear and the other nonlinear, we show the existence of our proposed spectral singularities. Then, we use an experimentally feasible realistic model based on coupled disk resonators and we identify the spectral singularity with directional sensitivity (SSDS). Our proposed SSDS might have applications in designing nonlinear sensors. Furthermore, it can provide a solution for the hole burning problem in pumped laser cavities.  
\end{abstract}


\maketitle
\section{Introduction}


While the study of the complex functions has a long history, in recent years they have attracted more attention due to their mathematical peculiar features and their applications specifically in optical systems. One of the unique features of complex potentials, for instance, is the existence of different types of singularities. Exceptional points are among such singularities with topological characteristics and arising when the Hamiltonian of the corresponding system becomes defective and the eigenvalues and their associated eigenstates coalesce. While direct physical identification of the exceptional points is difficult, their strong influence on the dynamics can be observed \cite{1,2,3,4}. Exceptional point singularities enable stop light \cite{5}, real-entire flat bands \cite{flatband}, unidirectional invisibility \cite{invis}, topological energy transfer \cite{topolo}, enhanced sensitivity \cite{lyang,mercedeh}, to name a few. 

Another type of singularities are spectral singularities related to the completeness of the continuous spectrum and can satisfy outgoing boundary conditions \cite{mostafa09}. In other words, spectral singularities do not correspond to square-integrable eigenfunctions. Within the scattering matrix formalism, such singularities identify the lasing threshold of cavities with gain \cite{7,8} where the cavity gain corresponds to a negative imaginary part of the refractive index. The notion of such spectral singularities extended to the semi-infinite lattices \cite{9}, nonlinear potentials \cite{10}, and nonreciprocal cavities in the presence of magnetic elements \cite{11}. In the latest one, the presence of a gyrotropic element together with the broken inversion symmetry in a 1D heterostructure results in asymmetric stationary inflection points where the group velocity of the wave vector in one direction becomes zero while in the opposite direction group velocity finds a finite and nonzero value. The asymmetric inflection points are the first modes that reach the lasing threshold and thus result in a robust unidirectional lasing. Unidirectional lasing modes are of interest in recent years due to their potential applications. Another approach to obtain directional emission is based on the strong asymmetric backscattering in the vicinity of an exceptional point \cite{ozdemir}. Nonlinear coupling between the clockwise and counterclockwise propagating waves in an ultrahigh-Q whispering-gallery micro-resonator can produce a chiral emission \cite{qcao}. Topological insulator lattices can generate directional lasing at the edge of the lattice in the presence of a gain mechanism \cite{banders}. Another type of spectral singularities with unidirectional response can obtain from the interplay of parity-time symmetry and Fano resonances \cite{khodamuni}. In such spectral singularities, without breaking the reciprocity, one is able to obtain a simultaneous unidirectional lasing and unidirectional reflection-less mode. For such a mode one side reflection tends to infinity, the other side reflection becomes zero, and the transmission coefficient remains finite. These singularities emerge from the resonance trapping and delay time associated with the reflected signal residing in the gain or loss part of the parity-time symmetric cavity \cite{khodamuni}.

In this paper, we introduce a new class of spectral singularities with sensitivity to the direction of excitation. Such spectral singularities do not generate directional lasing, however, their source of the emission comes from a specific direction. In one dimension, our spectral singularities pick up fluctuation coming from one direction and thus lasing emission is activated by the fluctuation from one side. Our proposed spectral singularities appear in systems with broken parity and in the presence of nonlinearity. While we are interested to mathematically prove the existence of such spectral singularities one might be able to find their application in directional sensing and suppression of spatial hole burning \cite{holeburn}.

This article is structured as follows. In Sec. \ref{sec2}, we will review the scattering formalism for an abstract model, namely a nonlinear $\delta$-function, and calculate the spectral singularities associated with it. In Sec. \ref{sec3}, we will construct a system composed of a linear and a nonlinear $\delta$-function with broken parity symmetry. We construct the scattering matrix and calculate the spectral singularities that appear in the system. Finally, in Sec.(\ref{secdisk}), we discuss the appearance of SSDS in a coupled disk resonator system composed of two coupled resonators, one linear and the other nonlinear that are coupled to a 1D transmission line. We will draw our conclusion in Sec. \ref{sec4}.

\section{Scattering formalism for a nonlinear potential and its spectral singularities} \label{sec2}
In this section, we will briefly review the basic method for treating the scattering properties of a nonlinear $\delta$-function potential and constructing the scattering matrix in a 1D nonlinear problem. Furthermore, we connect the scattering matrix $S$ to the spectral singularities. Finally, in this section, we calculate the spectral singularities associated with our setup. We show that the spectral singularities in a single nonlinear $\delta$-function potential are reciprocal. In other words, at the spectral singularity, all the reflection and transmission coefficient tends to infinity with the same slope. This occurs because the system preserves the parity symmetry even it has a nonlinear component.

Let us consider a system that its permittivity is given by $\epsilon(x)=n_0+(n+\chi |E(x)|^2)\delta(x) $ where $\delta(x)$ is the Dirac delta function and $E$ is the electric field. In this permittivity $n_0$ is the background index of refraction (we assume it is equal to one), $n=n_{r}+in_{i}$ and $\chi=\chi_{r}+\chi_i$, where $n_{r,i}\in \Re$, and $\chi_{i,r}\in \Re$, are the linear and nonlinear complex perturbations to the background permittivity. Such permittivity can be realized with very thin layers of coated materials.  In this arrangement, a time-harmonic electric field of frequency $\omega$ obeys the 1D Helmholtz equation:
\begin{equation}
\frac{d^2}{dx^2}E(x)+\frac{\omega^2}{c^2}\epsilon(x) E(x)=0.
\label{eq1}
\end{equation}
In Eq.(\ref{eq1}), $c$ is the speed of light in the vacuum. On the left and right side of the delta function potential, Eq. (\ref{eq1}) admits the solution $E^{-}(x)=E_f^{-}\exp(ikx)+E_b^{-}\exp(-ikx)$ for $x\leq 0$ and $E^{+}(x)=E_f^{+}\exp(ikx)+E_b^{+}\exp(-ikx)$ for $x\geq 0$ where the wave vector $k=n_0\omega/c$. 

Although the problem in hand is nonlinear, still one can use the $S$-matrix formalism to treat the scattering properties of it. More precisely, the amplitudes of the ingoing and outgoing propagating waves outside the scattering domain, namely ($E_{f}^-,E_{b}^+$) and ($E_{b}^-,E_{f}^+$) respectively,
are related through the nonlinear $2\times 2$ scattering matrix $S$:
\begin{equation}
\left(\begin{array}{c}
E_{b}^-\\
E_{f}^+
\end{array}\right)=
S
\left(\begin{array}{c}
E_{f}^-\\
E_{b}^+
\end{array}\right).
\label{eqs0}
\end{equation}
In the above formulation, the elements of the scattering matrix are related to the transmission and reflection amplitudes for the left and right incidents, namely
\begin{equation}
S=\left(\begin{array}{cc}
S_{11}&S_{12}\\
S_{21}&S_{22}
\end{array}\right)=\left(\begin{array}{cc}
r_L&t_R\\
t_L&r_R
\end{array}\right).
\label{eqS}
\end{equation}
In the Eq.(\ref{eqS}), $r_{L(R)}$ and $t_{L(R)}$ are the reflection and transmission amplitude for the left, $L$ (right, $R$) incident waves, respectively. In nonreciprocal structures, the elements of the $S$ matrix in Eq.(\ref{eqS}) might be different from each other. In linear systems and in the absent of the magnetic field or spatiotemporal modulation reciprocity states that $S_{12}=S_{21}$ \cite{recip}. 

In the reciprocal systems and at the real frequencies, spectral singularities have been identified as the poles of the scattering matrix which is equivalent to the blowup of the transmission and reflections amplitudes. This defines the lasing points where for no in-going field there is an outgoing field solution in Eq.(\ref{eqs0}). Generally, for an open scattering system with no embedded gain medium the poles of the scattering matrix occur at complex frequencies with a negative imaginary part. By introducing gain in the system, the absolute value of the imaginary part of such frequencies decreases until at the lasing threshold one of the poles reaches the real axes. The real part of the frequency of this pole describes the frequency of the first lasing mode. In some systems such as ring-lasers several modes can reach the lasing threshold at the same time, the so-called multi-mode lasing, which is an undesired phenomenon as it distributes gain power between several modes and reduces the lasing power at the desired mode. 

Recently new spectral singularities have been introduced where result in unidirectional lasing mode. Specifically in the presence of reciprocity, at these spectral singularities only one reflection, namely one of the diagonal terms of the $S$ matrix in Eq.(\ref{eqS}), tends to infinity and other elements of the $S$ matrix remain finite\cite{khodamuni}. In the nonreciprocal system and in the presence of a magnetic field it has been shown that such spectral singularities result in the infinite transmission and reflection in one side of the structure. In this case only one row of the $S$ matrix in Eq.(\ref{eqS}) tends to infinity while the other row remains finite which is equivalent to having either left transmission coefficient $T_L=|t_L|^2\to \infty$ and right reflection coefficient  $R_R=|r_R|^2\to \infty$ or right transmission coefficient $T_R=|t_R|^2\to \infty$ and left reflection coefficient $R_L=|r_L|^2\to \infty$ \cite{11}.

The spectral singularities that we are looking for them here, have different characteristics from the aforementioned ones. Specifically, while at these spectral singularities only specific elements of the scattering matrix in Eq. (\ref{eqS}) tends to infinity, they do not cause any directional lasing. At these singularities only one reflection coefficient $R_{L(R)}$ and one transmission coefficient $T_{L(R)}$ tend to infinity while the other reflection coefficient, $R_{R(L)}$, and transmission coefficient, $T_{R(L)}$, remain finite. In other words, one column of the scattering matrix $S$ in Eq.(\ref{eqS}) tends to infinity while the other column remains finite. Thus, one can claim that these spectral singularities are sensitive to the direction of incident fluctuations. In Sec.{\ref{sec3}} we show that a parity broken nonlinear scattering system enables us to realize such spectral singularities with directional sensitivity (SSDS).

Let us go back to the parity symmetric single nonlinear $\delta$-function problem. The transmission and reflection amplitudes for left and right incident waves can be obtained from the boundary conditions 
\begin{equation}
E_b^+=0 \quad \text{and} \quad E_f^-=0
\label{eqb}
\end{equation}
respectively, and are defined as 
\begin{equation}
\begin{array}{cc}
t_L\equiv {E_f^{+}\over E_f^{-}}, & r_L\equiv 
{E_b^{-}\over E_f^{-}}\\
t_R\equiv {E_b^{-}\over E_b^{+}}, & r_R\equiv {E_f^{+}\over E_b^{+}}.
\end{array}
\label{eqtr}
\end{equation}

For the left incident field continuity of the field at the $\delta$-function potential, namely
\begin{equation}
E^-(x)|_{x=0}=E^+(x)|_{x=0}
\label{eqc}
\end{equation} and the discontinuity of it at $x=0$, namely
\begin{equation}
\frac{dE^+}{dx}|_{x=0}-\frac{dE^-}{dx}|_{x=0}=-k^2(n+\chi |E^+(0)|^2)E^+(0),
\label{eqld}
\end{equation}
provide us with two conditions to find $E^{-}_f$ and $E^{-}_b$ as a function of $E^{+}_f$ (here, we normalize it to one) for the left incident beam. Notice that inserting the first boundary condition in Eq.(\ref{eqb}) into the Eq.(\ref{eqld}) reduces it to 
\begin{equation}
\frac{dE^+}{dx}|_{x=0}-\frac{dE^-}{dx}|_{x=0}=-k^2(n+\chi)E^+(0),
\label{eqld1}
\end{equation}
which means that in this specific case nonlinearity acts similar to a linear potential and the field intensity does not play any role even the potential at hand is nonlinear. This is a peculiar feature of $\delta$-function and does not hold in nonlinear slab potentials.
Solving Eqs.(\ref{eqc}) and(\ref{eqld1}) simultaneously results in the transmission, $S_{21}\equiv t_L$, and reflection, $S_{11}\equiv r_L$, amplitudes for the left incident wave:
\begin{equation}
t_L=\frac{2 i}{k (n+\chi )+2 i}, \quad r_L=-\frac{k (n+\chi )}{k (n+\chi )+2 i}.
\label{sdli}
\end{equation}

For the right incident field Eq.(\ref{eqc}) remains the same while Eq.(\ref{eqld1}) modifies to 
\begin{equation}
\frac{dE^+}{dx}|_{x=0}-\frac{dE^-}{dx}|_{x=0}=-k^2(n+\chi)E^-(0),
\label{eqrd1}
\end{equation}
where we have used the second boundary condition given in Eq.(\ref{eqb}). Normalizing the $E^{-}_b$ to one and following similar steps as the left incident wave, results in the transmission $S_{12}\equiv t_R$ and reflection $S_{22}\equiv r_R$ amplitudes for the right incident wave and comparing them with their corresponding left incident ones shows that
\begin{equation}
t_R=t_L, \quad r_R=r_L.
\label{sdri}
\end{equation}
Equation (\ref{sdri}) is a known result where a nonlinear medium with parity symmetry does not lead to any asymmetric transport\cite{casati}. Using Eqs. (\ref{eqS}, \ref{sdli} and \ref{sdri}) we can construct the scattering matrix $S$. 

As mentioned earlier the poles of this scattering matrix at real frequencies identify the spectral singularities or the lasing modes. From Eq.(\ref{sdli}) and Eq.(\ref{sdri}) we observe that the spectral singularities occur for pure imaginary values of $n$ and $\chi$ and is given by $k=-\frac{2}{n_i+\chi_i}$. Furthermore, with our choice of coordinate to have the outgoing fields, wave vector $k$ must be a positive variable and thus $n_i+\chi_i<0$ meaning that although $\delta$-function can contain partial loss either in its linear $n_i>0$ or nonlinear part$\chi_i>0$, it must provide a net gain coming from its nonlinear $\chi_i<0$ or linear $n_i<0$ part of it. This conclusion proves our previous discussion where we mentioned that one needs to incorporate a sufficiently strong gain medium to reach the lasing threshold. 
\begin{figure}
	\includegraphics[width=1\linewidth, angle=0]{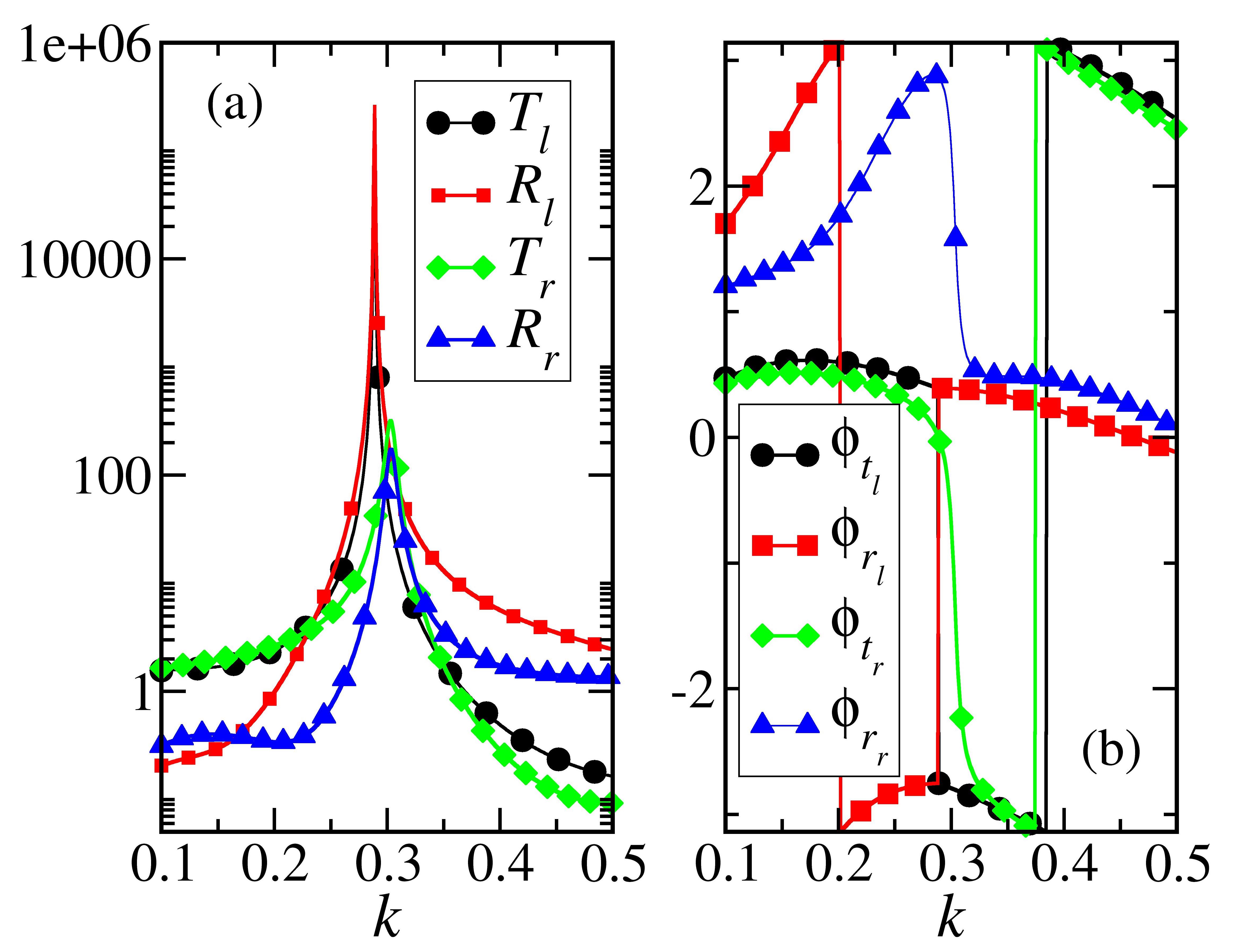}
	\caption{(Color online) Left spectral singularity with directional singularity. (a) Transmissions and reflections curves versus wave-vector $k$ associated with a system composed of two $\delta$-function, one linear with permittivity $n_1=2-3.4641i$ placed at $x=-2.7207$	(in the unite of $k^{-1}$) and the other with $n_2=3-3.4641i$ and $\chi=1$ placed at $x=2.7207$	(in the unite of $k^{-1}$). Transmission left and reflection left are diverging at $k\approx0.288675$ while transmission and reflection for the right incident beam remain finite. (b) The phase associated with the $t_{i},r_{i}$, $i=l,r$. At the SSDS there a $\pi$ shift occurs for the left reflection and transmission while the right reflection and transmission curves are smooth. \label{figdeltaleft}}
	
\end{figure}

\section{Spectral singularities with directional sensitivity} 
\label{sec3}
Armed with the method of calculating the spectral singularities given in Sec.\ref{sec2} in this section we calculate the spectral singularity associated with a parity symmetry broken nonlinear system. To demonstrate the existence of the spectral singularities with directional sensitivity we consider a system that its permittivity is given by $\epsilon(x)=n_0+n_1\delta(x+a)+(n_2+\chi|E(x)|^2)\delta(x-a)$. In this permittivity $n_{1}=n_{1r}+in_{1i}$, $n_{2}=n_{2r}+in_{2i}$ and $\chi=\chi_{r}+\chi_i$, where $n_{1r} \text{, }n_{2r}\text{, }n_{1i}\text{, }n_{2i}\text{, } \text{and } \chi_{i,r}\in \Re$ are the linear and nonlinear complex perturbations to the background permittivity. First $\delta$-function which is given by the perturbed refractive index $n_1$ is placed at $x=-a$ and the second one with amplitude $n_2+\chi|E(x)|^2$ is placed at $x=a$.

Similar to the previous scattering problem a time-harmonic electric field of frequency $\omega$ obeys the 1D Helmholtz equation which is given in Eq.(\ref{eq1}) with the following general solution
\begin{equation}
\begin{array}{cc}
E^{-}(x)=E_f^{-}\exp(ikx)+E_b^{-}\exp(-ikx) & x\leq-a\\
E^{m}(x)=E_f^{m}\exp(ikx)+E_b^{m}\exp(-ikx) & -a\leq x\leq a\\
E^{+}(x)=E_f^{+}\exp(ikx)+E_b^{+}\exp(-ikx)& x\geq a
\end{array}
\nonumber
\end{equation}
with the wave vector $k=n_0\omega/c>0$.

Following similar steps as the one discussed in Sec.(\ref{sec2}), first we calculate the scattering properties of the medium for the left incident field. In this case, we assume $E^+_f=1,E^+_b=0$. Using the continuity of the field and discontinuity of its derivative at $x=a$ and $x=-a$, one can show that $E^m_f=1-\frac{1}{2} i k (n_2+\chi)$, $E^m_b=\frac{1}{2} i k e^{2 i a k} (n_2+\chi )$ and the transmission and reflection amplitudes for a left incident beam are given by

\begin{equation}
\begin{split}
t_l&=\frac{4}{k^2 n_1 \zeta\eta-2 i k \left(n_1+\eta \right)+4}\\
r_l&=\frac{e^{-2 i a k} \left(k n_1 \left(k \eta+2 i\right)-k e^{4 i a k} \left(k n_1-2 i\right) \eta\right)}{k^2 n_1 \zeta \eta-2 i k \left(n_1+\eta\right)+4}.
\end{split}
\label{2nonleft}
\end{equation} 
In the above equations the new parameters are defined as $\zeta \equiv\exp({4 i a k})-1$ and $\eta=n_2+\chi$.

To calculate the transmission and reflection amplitude of the right incident field, on the other hand, we assume that $E^-_f=0,E^-_b=1$. By imposing the boundary conditions at each delta function, namely the continuity of the field and discontinuity of its derivative at $x=\pm a$, we find that  $E^m_f=\frac{1}{2} i k n_1 e^{2 i a k}$, $E^m_b=1-\frac{i k n_1}{2}$ while the transmission and reflection amplitude for a right incident beam are given by
\begin{widetext}
	\begin{equation}
	\begin{split}
	t_r&=\frac{16 e^{4 i a k}}{k^3 \chi \left| n_1\right|^2 \zeta^2 \left(4 i-k n_1 \zeta\right)+2 e^{4 i a k} \left( k^2 n_1 \zeta \left(2n_2+4 \chi +i k^2 n_1 \chi  \zeta\right)+2 k^2 \chi n_1^*-4 i k\left(n_1+n_2\right)-4 i k \chi +8\right)-4 k^2 \chi  n_1^*}\\
	r_r&=\frac{ k^2 \chi  \zeta^2 \left| n_1\right|^2 \left(4 i-k n_1 \zeta\right)+4 e^{4 i a k} \left(k \chi  n_1^*-k n_1 \eta-2 i \eta \right)+4 n_1 e^{8 i a k} \left( kn_2-2 i k^2 n_1 \chi  \sin ^2(2 a k)+2 k \chi -2 i\right)-4 k \chi  n_1^*}{k \chi  e^{2 i a k} \zeta n_1^* \left(k n_1 \zeta-2 i\right){}^2+2 e^{6 i a k} \left( n_1 \left(-i k^2 n_1 \chi  \zeta^2-2 k e^{4 i a k} \left(\eta+ \chi \right)+2 k (\eta+ \chi) +4 i\right)+4 i  \eta -8k^{-1}\right)}
	\end{split}
	\label{2nonright}
	\end{equation} 
\end{widetext}

It is clear that $r_l\neq r_r$ and $t_l\neq t_r$ where the first appears due to the broken parity symmetry and the second occurs due to the coexistence of the nonlinearity and broken parity symmetry\cite{casati}. 

As discussed before, to identify the spectral singularities we are interested in the zeros of the denominators in Eq.(\ref{2nonleft}) and Eq.(\ref{2nonright}). The clear difference between the denominators in Eqs.(\ref{2nonleft},\ref{2nonright}) makes it possible to search for the poles of the scattering matrix $S$ such that only one column remains finite while the other column diverges and thus the system finds itself at the SSDS mode.

In general finding, an analytical solution for an SSDS of the above system is not possible. However, one can find the SSDS points for the left and right incident fields numerically. In an attempt to find the SSDS for the left and right incident field let us assume that $a=\frac{\pi}{4k^\star}$ where $k^\star$ is the wave-vector for which the SSDS mode occurs. This choice does not affect the physics of the problem and it only saves us from doing difficult numerical tasks. For the left incident field the zeros of the denominators in Eq.(\ref{2nonleft}) occur at 
\begin{equation}
k^\star=\pm\frac{4}{\sqrt{6 n_1 \eta-\eta^2-n_1^2}\pm i( n_1+\eta) }.
\label{kstarleft}
\end{equation}
For arbitrary values of $n_1,n_2,\chi$ the wave-vectors in Eq.(\ref{kstarleft}) are complex. However, the SSDS wave-vectors must be positive and real in order to identify a lasing point. Thus, if there exist $n_1,n_2$, and $\chi$ such that make the $k^\star$ in the Eq.(\ref{kstarleft}) to be real then we can only accept the positive solution. As an example lets assume that $n_1=2+i\gamma$, $n_2=3+i\gamma$, and $\chi=3$. A numerical search shows that when $a=2.7207$ for $\gamma=-3.4641$ the SSDS wave-vector for the left incident field becomes real with value $k^\star=0.288675$. As a result, we expect to have a divergence in the left transmission and left reflection while the right transmission and right reflection remain finite. In figure (\ref{figdeltaleft}a) we have plotted the transmissions and reflections for the above $\delta$-functions. In an agreement with our analytical and numerical predictions the left transmission and reflection amplitudes diverge at $k^\star=0.288675$ while the $t_r$ and $r_r$ remain finite. The right transmission and reflection show a strong amplification at $k_0\approx 0.3>k^\star$ which is different from a diverging behavior. The difference between the wave-vectors associated with the right amplification and left SSDS, $\Delta k\equiv|k_0-k^\star|$ is proportional to $\chi$ and tends to zero when the nonlinearity coefficient $\chi$ decreases to zero. It is known that at the spectral singularity the phase must obtain a $\pi$ shift \cite{khodamuni}. Figure (\ref{figdeltaleft}b) depicts the phase of the transmission and reflection amplitudes. We see that $\pi$ jump occurs at the $k^\star$ for the phases of the left transmission, $\phi_{t_l}$ and left reflection $\phi_{r_l}$ while the curves associated with the right reflection $\phi_{r_r}$ and transmission $\phi_{t_r}$ are smooth curves.

\begin{figure}
	\includegraphics[width=1\linewidth, angle=0]{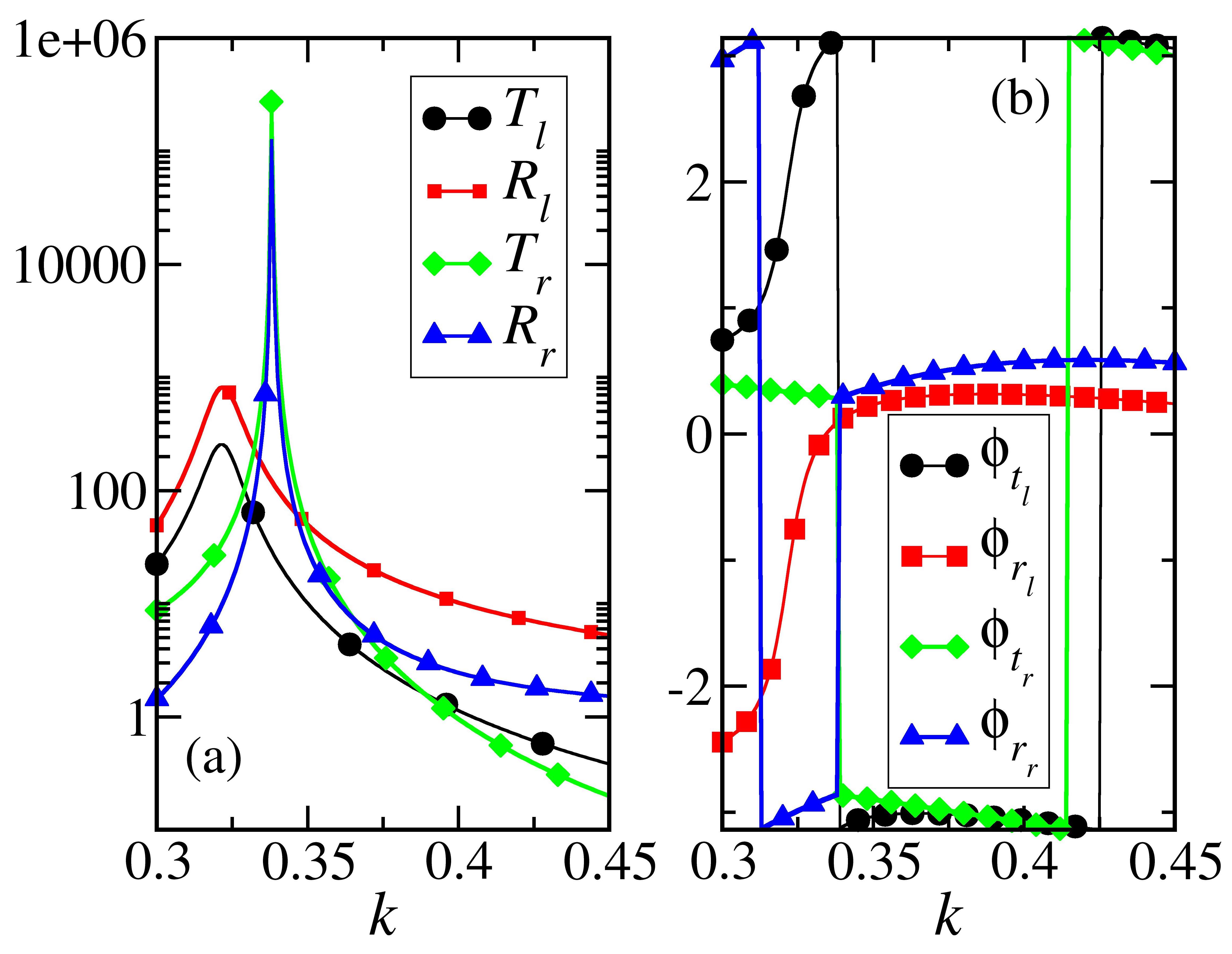}
	\caption{(Color online) Right spectral singularity with directional singularity. (a,b) Similar to the one presented in Fig.(\ref{figdeltaleft}) with delta functions placed at $x\approx \pm 2.32$ with $n_1=2-2.95712i$, $n_2=3-2.95712i$, and $\chi=3$. 
		\label{figdeltaright}}
	
\end{figure}
Unfortunately, there is no close form for the right SSDS wave-vector. However, still one can numerically locate those wave-vectors that diverge the amplitudes gave in Eq.(\ref{2nonright}). For example when the two delta function are apart by a distance equal to $a=4.64503$ and $n_1=2-2.95712i$, $n_2=3-2.95712i$ and $\chi=3$, the SSDS wave-vector for the right incident field becomes $k^\star=0.338167$. This numerical prediction is in agreement with our calculated $t_{i}$ and $r_{i}$, with $i=l,r$ and their corresponding phases in figures (\ref{figdeltaright}a) and (\ref{figdeltaright}b). 

\section{Spectral singularity with directional sensitivity in coupled micro-resonators}\label{secdisk}
\begin{figure}
	\includegraphics[width=1\linewidth, angle=0]{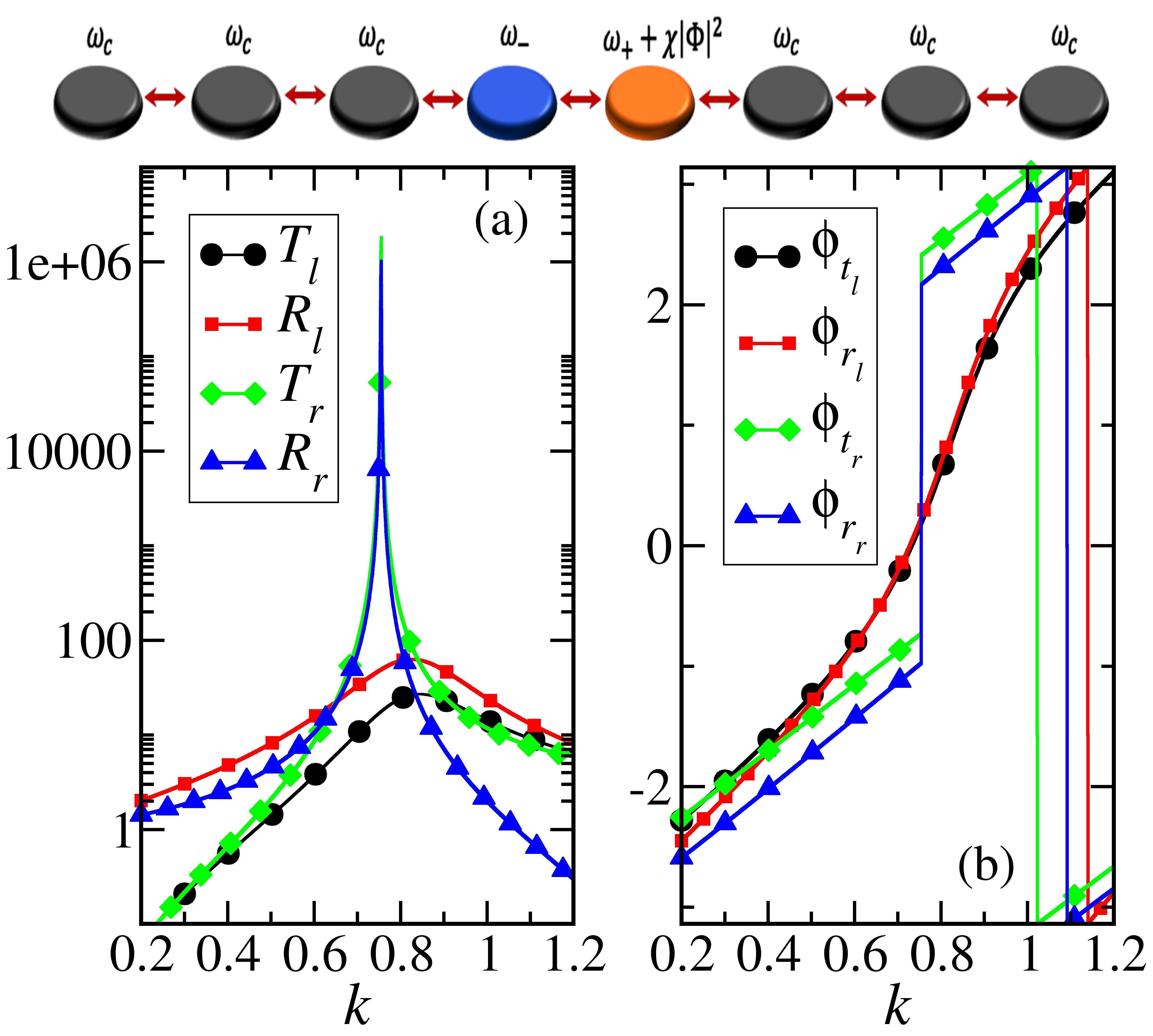}
	\caption{(Color online) (a) Transmission and reflection coefficient of the coupled disk resonators depicted in the upper inset. The resonance frequency of the disks at the transmission line are $\omega_c=0$, resonance frequency of the linear (blue) disk with gain is $\omega+{\_}=0.503461i$ while linear part of the resonance frequency of the nonlinear disk is $\omega_+=2\omega+{\_}$ and its Kerr coefficient is $\chi=1$. Transmission right and reflection right are diverging at $k^\star=0.754559$ (in the unite of couplings) indicating the existence of right SSDS mode. (b) corresponding phases. We observe the $\pi$ phase shift at the spectral singularity. \label{disk}}
\end{figure}

In the previous section, we used an abstract model to show the existence of SSDS modes. In this section, we identify the SSDS modes in an experimentally feasible system, namely an array of coupled disk resonators depicted schematically in the upper inset of the Fig.(\ref{disk}). The array composed of two distinct disk resonators, one with a linear resonance frequency and the other with a nonlinear resonance frequency, embedded in an array of coupled disk resonators playing the role of the transmission line. In this system, each disk supports two degenerate modes: a clockwise $a^+$ and a counterclockwise $a^-$. Using the coupled mode theory, we express the dynamics of the total field amplitudes $\Phi=a^++a^-$ in each disk \cite{lev, note}
\begin{equation}
\begin{split}
i\dot\Phi_n&=-\Phi_{n-1}-\Phi_{n+1}+\omega_c \Phi_n\quad (|n|\ge 2)\\
i\dot\Phi_n&=-\delta_{n\pm 1,2}\Phi_{n\pm 1}-\delta_{n,\pm 1}\Phi_{\pm}+\omega_c \Phi_n \quad (n=\pm 1)\\
i\dot\Phi_+&=-\Phi_{1}-\Phi_{\_}+(\omega_+ +\chi|\Phi_+|^2) \Phi_+\quad \\
i\dot\Phi_{\_}&=-\Phi_{-1}-\Phi_{+}+\omega_{\_}  \Phi_{\_}\quad \\
\end{split}\label{eqr0}
\end{equation}
where we used Kronecker delta notation 
\begin{equation}
\delta_{i,j}=\begin{cases}
1       & \quad \text{if } i =j\\
0  & \quad \text{if } i\neq j
\end{cases}.
\end{equation}
Furthermore, $\Phi_n$, and $\Phi_{\pm}$ are the total field amplitude at the disk $n$, and nonlinear (linear) cavity with subindex $+(-)$, respectively. Note that we have assumed all the couplings are equal (normalized to one) and the nonlinear disk has a Kerr type nonlinearity. The resonance frequency of the disks in the chain is denoted by $\omega_c$ which without lose of generality we assume it is equal to zero. Moreover, the resonance of the nonlinear disk is $(\omega_+ +\chi|\Phi_+|^2)$ while the resonance of the linear disk with field amplitude $\phi_-$ is given by $\omega_-$. The chain supports the dispersion relation $\omega=-2 \cos (k)$, with $-\pi\leq k \leq \pi$. In the elastic scattering process for which $\Phi=\psi e^{-i\omega t}$, the stationary modal amplitudes of the system have the asymptotic behavior $\psi_n=F_l e^{i k(n+1)}+B_l e^{-i k (n+1)}$ for $n\leq -1$ and $\psi_n=F_r e^{i k(n-1)}+B_r e^{-i k (n-1)}$ for $n\geq 1$, respectively.

For the left incident field where $B_r=0$ and $F_r=1$ it is easy to show that 
\begin{equation}
\begin{split}
t_l&=\frac{e^{3 i k}-e^{5 i q}}{1+e^{i k} \left(\omega _{\_}+\beta\right)+e^{2 i k} \left(\beta\omega _{\_} -1\right)}\\
r_l&=\frac{-e^{3 i k} \left(\omega _{\_}+e^{i k} \left(\omega _{\_}+e^{i k}\right) \beta\right)}{1+e^{i k} \left(\omega _{\_}+\beta\right)+e^{2 i k} \left(\beta\omega _{\_} -1\right)}
\end{split}
\label{diskleft}
\end{equation}
and the left SSDS mode wave-vector is given by
\begin{equation}
k^\star=-i \ln \left(\frac{-2}{\omega _{\_}+\beta\pm\sqrt{\left(\beta-\omega _{\_}\right){}^2+4}}\right)
\label{pp}
\end{equation}
where $\beta=\chi+\omega_+$. Assuming $B_l=1$ and $F_l=0$, transmission right and reflection right can also be calculated 
\begin{widetext}
	\begin{equation}
	\begin{split}
	t_r&=\frac{e^{3 i k}-e^{5 i k}}{\chi  \omega _{\_}{}^* \left(1+e^{i k} \omega _{\_}\right){}^2+e^{3 i k} \chi  \omega _{\_}^2+e^{i k} \left(\omega _{\_}+\beta\right)+e^{2 i k} \left(\omega _{\_} \left(2 \chi +\omega _+\right)-1\right)+1}\\
	r_r&=\frac{e^{4 i k} \left(1-e^{-i k} \left(1+e^{i k} \omega _{\_}\right) \left(\chi  \left| \omega _{\_}\right| {}^2+2 e^{-i k} \chi  \Re\left(\omega _{\_}\right)+i \sin (k) \left(2 \chi  \omega _{\_}+1\right)+\cos (k)+\beta\right)\right)}{e^{i k} \left(2 \chi  \left| \omega _{\_}\right| {}^2+\omega _{\_}+\beta\right)+e^{2 i k} \left(\omega _{\_} \left(\chi  \left(\left| \omega _{\_}\right| {}^2+2\right)+\omega _+\right)-1\right)-i \chi  \Im\left(\omega _{\_}\right)+e^{3 i k} \chi  \omega _{\_}^2+\chi  \Re\left(\omega _{\_}\right)+1}
	\end{split}
	\label{diskright}
	\end{equation} 
\end{widetext}

In Eq.(\ref{diskright}), $\Re(\omega_{\_})$ and $\Im(\omega_{\_})$ are the real and imaginary part of the $\omega_{\_}$ respectively. The wave-vector for a right SSDS has a very long closed from and thus we do not report it here. Similar to the nonlinear delta function system one can numerically search for $\omega_{\pm}$ and $\chi$ such that make the $k^\star$, associated with a left SSDS (given in Eq.{\ref{pp}}) or right SSDS (not shown here), positive and real. For instance, the left SSDS occurs at $k^\star=0.950497$ (normalized to the unit of the couplings) when $\chi=1, \omega_{\_}=0.641322i$ and $\omega_+=2\omega_{\_}$ while the right SSDS occurs at $k^\star=0.754559$ when $\chi=1, \omega_{\_}=0.503461i$ and $\omega_+=2\omega_{\_}$. Thus, for this system the singularity mode (lasing threshold) appears first for the right fluctuations as a right SSDS is triggered by lower gain value. After that point the system becomes nonlinear and our treatment is not anymore valid. Thus, $k^\star=0.950497$ is not a physical solution. In figure (\ref{disk}) we reported the transmissions and reflections and their corresponding phases for $\chi=1, \omega_{\_}=0.503461$ and $\omega_+=2\omega_{\_}$. 
\section{Conclusion} \label{sec4}

In conclusion, we have developed scattering matrix formalism for nonlinear systems and then using an abstract model of two delta functions, one nonlinear and one linear, we have shown that a nonlinear system with broken parity symmetry can have peculiar spectral singularities with sensitivity to the direction of excitation, the so-called SSDS. We discussed the possibility of getting left or right SSDS depending on the distance between the two delta functions. Furthermore, we found the right SSDS in an experimentally feasible system, namely coupled disk resonators. While we have used a parity symmetry broken system to show the existence of SSDS, such singularities might be found in nonlinear systems with any broken symmetry that changes $\vec k\to -\vec k$. Our proposed spectral singularity might have application in directional sensing and can be considered as a solution for hole burning problems in laser cavities.

\begin{acknowledgments}
	H.R acknowledges the support by the Army Research Office Grant No. W911NF-20-1-0276 . The views and conclusions contained in this document are those of the authors and should not be interpreted as representing the official policies, either expressed or implied, of the Army Research Office or the U.S. Government. The U.S. Government is authorized to reproduce and distribute reprints for Government purposes notwithstanding any copyright notation herein. 
	
\end{acknowledgments}




\begin{thebibliography}{99}
	\bibitem{1} C. Dembowski, H.-D. Gr\"{a}f, H. L. Harney, A. Heine, W. D. Heiss, H. Rehfeld, and A. Richter, Phys. Rev. Lett. 86, 787 (2001); C. Dembowski, B. Dietz, H.-D. Gr\"{a}f, H. L. Harney, A. Heine, W. D. Heiss, and A. Richter, Phys. Rev. E 69, 056216 (2004); J. Schindler, Z. Lin, J. M. Lee, H. Ramezani, F. M. Ellis, and T. Kottos, Journal of Physics A: Mathematical and Theoretical 45, 444029 (2012).
	\bibitem{2} H. Ramezani, J. Schindler, F. M. Ellis, U. Gunther, and T. Kottos, Phys. Rev. A 85, 062122 (2012)
	
	\bibitem{3} X. Zhu, H. Ramezani, C. Shi, J. Zhu, and X. Zhang, Phys. Rev. X 4, 031042 (2014).
	
	\bibitem{4} F. Quijandría, U. Naether, Ş. K. Özdemir, F. Nori, and D. Zueco, “PT-Symmetric Circuit QED,” Phys. Rev. A 97, 053846 (2018).
	
	\bibitem{5}T. Goldzak, A. A. Mailybaev, and N. Moiseyev, Phys. Rev. Lett. 120, 013901 (2018).
	\bibitem{flatband} H. Ramezani, Physical review A 96 (1), 011802, (2017)
	\bibitem{invis} Z. Lin, et al. Phys. Rev. Lett. 106 (21), 213901 (2011)
	\bibitem{topolo} H. Xu, D. Mason, Luyao Jiang, J. G. E. Harris, Nature volume 537, (2016)
	\bibitem{lyang} W. Chen, \c{S}. K. \"{O}zdemir, G. Zhao, J. Wiersig, L. Yang, Nature, 548 (2017)
	\bibitem{mercedeh} H. Hodaei, et al., Nature 548 (2017)
	\bibitem{mostafa09} A. Mostafazadeh, Phys. Rev. Lett. 102, 220402 (2009).
	\bibitem{7} K. M. Farham et al., Europhys. Lett. 49, 48 (2000); M. Chitsazi et al., Phys. Rev. A 89, 043842 (2014); M. Brandstetter et al., Nat. Com. 5, 4034 (2014).
	
	\bibitem{8} For a review on scattering in 1D non-hermitian potential see J.G. Muga, J.P. Palaob, B. Navarroa, and I.L. Egusquiza, Phys. Rep. 395 (2004).
	\bibitem{9} S. Longhi, Phys. Rev. B 80, 165125 (2009).
	
	\bibitem{10} A. Mostafazadeh, Phys. Rev. Lett. 110, 260402 (2013); A. Mostafazadeh, Phys, Rev A, 87, 063838 (2013).
	
	\bibitem{11} H. Ramezani, S. Kalish, I. Vitebskiy, and T. Kottos, Phys. Rev. Lett. 112, 043904 (2014).
	\bibitem{ozdemir} B. Peng, et al. PNAS, 113 (25) 6845, 2016. 
	\bibitem{qcao} Q. Cao, et. al. Phys. Rev. Lett. 118, 033901 (2017)
	\bibitem{banders} M. A. Bandres, et al., Science, 359, 6381, p:4005, (2018)
	\bibitem{khodamuni} H. Ramezani, et al. , Phys. Rev. Lett. 113, 263905, (2014)
	\bibitem{holeburn} K. Schuhmann, K. Kirch, F. Nez, R. Pohl, G. Wichmann, and A. Antognini, Appl. Opt. 57, 2900-2908  (2018)
	\bibitem{recip} R. J. Potton, Rep. Prog. Phys. 67, 717 (2004)
	\bibitem{mostafaar} Ali Mostafazadeh, arXiv:1806.02610, (2018)
	\bibitem{CDV07} F. Cannata, J.-P. Dedonder and A. Ventura, Ann. of Phys. {\bf 322}, 397 (2007).
	\bibitem{casati} Stefano Lepri and Giulio Casati, Phys. Rev. Lett. 106, 164101 (2011)
	\bibitem{lev} M. F. Yanik and S. Fan, Phys. Rev. Lett. 92, 083901 (2004);
	H. Ramezani, T. Kottos, V. Shuvayev, and L. Deych, Phys.
	Rev. A 83, 053839 (2011).
	\bibitem{note} One can write the same set of equations for the subtraction
	of the clockwise and counterclockwise modes where all the couplings change their sign and the dispersion relation becomes $\omega=\omega_c+2 \cos (q)$. We carried out the same analysis and obtained the same results.
	
	
	
\end{thebibliography}
\end{document}